\documentclass{aa}  
\usepackage {graphicx}
\usepackage {times}
\usepackage {natbib}

\begin{document}

   \title{Environmental dependence of 8$\mu$m luminosity functions of galaxies at z$\sim$0.8}

   \subtitle{Comparison between RXJ1716.4$+$6708 and the AKARI NEP deep field.\thanks{This research is based on the observations with AKARI, a JAXA project with the participation of ESA.},\thanks{Based  on data collected at Subaru Telescope, which is operated by the National Astronomical Observatory of Japan.}}
\titlerunning{Environemental dependence of 8$\mu$m luminosity functions of galaxies at z$\sim$0.8}

   \author{Tomotsugu Goto
          \inst{1,2}\fnmsep\thanks{JSPS SPD fellow},
	  Yusei Koyama \inst{3},
	  T.Wada \inst{4},	
	  C.Pearson\inst{5,6,7},	  
	  H.Matsuhara\inst{4},
          T.Takagi\inst{4},	 
	  H. Shim \inst{8},
	  M.Im\inst{8},
          M.G.Lee\inst{8},
	  H.Inami\inst{4,9,10},
	  M.Malkan\inst{11},
	  S.Okamura\inst{3},	  
	  T.T.Takeuchi\inst{12},	
          S.Serjeant\inst{7},
	  T.Kodama\inst{2},	  
	  T.Nakagawa\inst{4},
	  S.Oyabu\inst{4},
	  Y.Ohyama\inst{13},
          H.M.Lee\inst{8},   
          N.Hwang\inst{2},
          H.Hanami\inst{14},
          K.Imai\inst{15},
and          T.Ishigaki\inst{16}
}

	  \institute{
	  Institute for Astronomy, University of Hawaii,	  2680 Woodlawn Drive, Honolulu, HI, 96822, USA\\
	       \email{tomo@ifa.hawaii.edu}
	       \and
	  National Astronomical Observatory, 2-21-1 Osawa, Mitaka, Tokyo, 181-8588,Japan
\and Department of Astronomy, School of Science, The University of Tokyo, Tokyo 113-0033, Japan
\and Institute of Space and Astronautical Science, Japan Aerospace Exploration Agency, 	     Sagamihara, Kanagawa 229-8510
\and Rutherford Appleton Laboratory, Chilton, Didcot, Oxfordshire OX11 0QX, UK
\and Department of Physics, University of Lethbridge, 4401 University Drive,Lethbridge, Alberta T1J 1B1, Canada
\and Astrophysics Group, Department of Physics,  The Open University, Milton Keynes, MK7 6AA, UK
\and Department of Physics \& Astronomy, FPRD, Seoul National University, Shillim-Dong, Kwanak-Gu, Seoul 151-742, Korea	
\and Spitzer Science Center, California Institute of Technology, Pasadena, CA 91125
\and Department of Astronomical Science,The Graduate University for Advanced Studies
\and Department of Physics and Astronomy, UCLA, Los Angeles, CA, 90095-1547 USA
\and Institute for Advanced Research, Nagoya University, Furo-cho, Chikusa-ku, Nagoya 464-8601
\and Academia Sinica, Institute of Astronomy and Astrophysics, Taiwan
\and Physics Section, Faculty of Humanities and Social Sciences, Iwate University, Morioka, 020-8550
\and  TOME R\&D Inc. Kawasaki, Kanagawa 213 0012, Japan
\and Asahikawa National College of Technology, 2-1-6 2-jo Shunkohdai, Asahikawa-shi, Hokkaido 071-8142 
	     }
   \date{Received September 15, 2009; accepted December 16, 2009}
   \authorrunning{Goto et al.}

 
  \abstract
{}
   {
 We aim to reveal environmental dependence of infrared luminosity functions (IR LFs) of galaxies at z$\sim$0.8 using the AKARI satellite. AKARI's wide field of view and unique mid-IR filters help us to construct restframe 8$\mu$m LFs directly without relying on SED models.
   }
   {
We construct restframe 8$\mu$m IR LFs in the cluster region RXJ1716.4$+$6708 at z=0.81, and compare them with a blank field using the AKARI North Ecliptic Pole deep field data at the same redshift. 
AKARI's wide field of view (10'$\times$10') is suitable to investigate wide range of galaxy environments. AKARI's 15$\mu$m filter is advantageous here since it directly probes restframe 8$\mu$m at $z\sim$0.8, without relying on a large extrapolation based on a SED fit, which was the largest uncertainty in previous work. 
}
   {
We have found that cluster IR LFs at restframe 8$\mu$m have a factor of 2.4 smaller $L^*$ and a steeper faint-end slope than that of the field. 
 Confirming this trend, we also found that faint-end slopes of the cluster LFs becomes flatter and flatter with decreasing local galaxy density.
 These changes in LFs cannot be explained by a simple infall of field galaxy population into a cluster. 
 Physics that can preferentially suppress IR luminous galaxies in high density regions is required to explain the observed results. 
}
   {}


  \maketitle

\section{Introduction}

It has been observed that galaxy properties change as a function of galaxy environment; 
the morphology-density relation reports that fraction of elliptical galaxies is larger at higher galaxy density \citep{2003MNRAS.346..601G};
the star formation rate (SFR) is higher in lower galaxy density \citep{2003ApJ...584..210G,2004AJ....128.2677T} .
However, despite accumulating observational evidence, we still do not fully understand the underlying physics governing environmental-dependent evolution of galaxies. 

Infrared (IR) emission of galaxies is an important probe of galaxy activity since at higher redshift, a significant fraction of star formation is obscured by dust \citep{2005A&A...440L..17T,GOTO_NEP_LF}.
Although there exist low-z cluster studies \citep{2006ApJ...639..827B,Shim2010,Tran2010}, not much attention has been paid to the infrared properties of high redshift cluster galaxies, mainly due to the lack of sensitivity in previous IR satellites such as ISO and IRAS. 
 Superb sensitivity of recently launched Spitzer and AKARI satellites can revolutionize the infrared view of environmental dependence of galaxy evolution. 

In this work, we compare restframe 8$\mu$m LFs between cluster and field regions at z=0.8 using data from the AKARI.
 Monochromatic restframe 8$\mu$m luminosity ($L_{8\mu m}$) is important since it is known to correlate well with the total IR luminosity \citep{2006MNRAS.370.1159B,2007ApJ...664..840H},  and hence, with the SFR of galaxies \citep{1998ARA&A..36..189K}. 
 This is especially true for star-forming galaxies because the rest-frame 8$\mu$m flux are dominated by prominent PAH features such as at 6.2, 7.7 and 8.6 $\mu$m  \citep{1990A&A...237..215D}.

%

Important advantages brought by the AKARI are as follows: 
(i) At z=0.8, AKARI's 15$\mu$m filter ($L15$) covers the redshifted restframe 8$\mu m$, thus we can estimate 8$\mu$m LFs without using a large extrapolation based on SED models, which were the largest uncertainty in previous work.
(ii) Large field of view of the AKARI's mid-IR camera (IRC, 10'$\times$10') allows us to study wider area including cluster outskirts, where important evolutionary mechanisms are suggested to be at work \citep{2004MNRAS.348..515G,2004MNRAS.354.1103K}.
 For example, passive spiral galaxies have been observed in such an environment \citep{2003PASJ...55..757G}.
  Unless otherwise stated, we adopt a cosmology with $(h,\Omega_m,\Omega_\Lambda) = (0.7,0.3,0.7)$ \citep{2009ApJS..180..330K}.

\section{Data \& Analysis}
\label{Data}

\begin{figure}
\begin{center}
\includegraphics[scale=0.6]{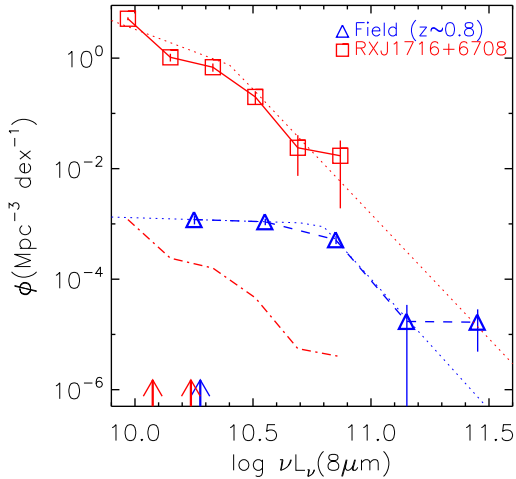}
\end{center}
\caption{
Restframe 8$\mu$m LFs of cluster RXJ1716.4$+$6708 at z=0.81 in the squares, and those of the AKARI NEP deep field in the triangles.
For  RXJ1716.4$+$6708, only photometric and spectroscopic cluster member galaxies are used. For the NEP deep field, galaxies with photo-z/specz in the range of $0.65<z<0.9$ are used. The dot-dashed lines are 8$\mu$m LFs of RXJ1716.4$+$6708, but scaled down for easier comparison. The thin dotted lines are the best-fit double power laws. Vertical arrows show the 5$\sigma$ flux limits of deep/shallow regions of the cluster (red) and the NEP deep field (blue) in terms of $L_{8\mu m}$ at z=0.81.
}\label{fig:8umlf}
\end{figure}

\begin{figure}
\begin{center}
\includegraphics[scale=0.6]{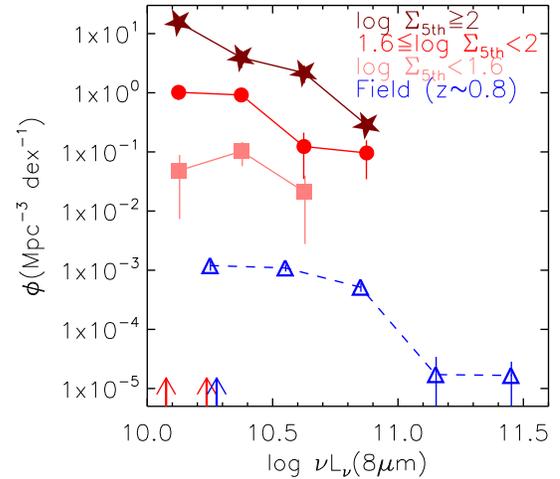}
\end{center}
\caption{ 
Restframe 8$\mu$m LFs of cluster RXJ1716.4$+$6708 at z=0.81, divided according to the local galaxy density ($\Sigma_{5th}$).
The stars, circles and squares are for galaxies with $log \Sigma_{5th}\geq 2 $, $1.6\leq log \Sigma_{5th}<2$, and $log \Sigma_{5th}< 1.6$, respectively.
}\label{fig:8umlf_density}
\end{figure}

\begin{small}
\begin{table*}
 \centering
  \caption{Best double power-law fit parameters for LFs}\label{tab:fit_parameters}
  \begin{tabular}{@{}ccclllcccc@{}}
  \hline
   Sample &  $L_{8\mu m}^*$ ($L_{\odot}$)&{\tiny  $\phi^*(\mathrm{Mpc^{-3} dex^{-1}})$} & $\alpha$ & $\beta$  \\ 
 \hline
 \hline
 NEP Deep field     & 6.1$\pm0.5\times 10^{10}$ &   0.0010$\pm$0.0003   & 1.1$\pm$0.3 &      5.7$\pm$1.2	\\
{\tiny RXJ1716.4$+$6708}    & 2.5$\pm0.1\times 10^{10}$ &   0.74$\pm$0.04&       2.6$\pm$0.1&      5.5$\pm$0.4\\
 \hline
\end{tabular}
\end{table*}
\end{small}

\subsection{LFs of cluster RXJ1716.4$+$6708}
The AKARI is a Japanese infrared satellite \citep{2007PASJ...59S.369M}, which has continuous filter coverage in the mid IR wavelengths  ($N2,N3,N4,S7,S9W,S11,L15,L18W$ and $L24$).
The AKARI has observed a massive galaxy cluster, RXJ1716.4$+$6708, in $N3, S7$ and $L15$ \citep{2008MNRAS.391.1758K}. 
 RXJ1716.4$+$6708 is at z=0.81 and has $\sigma=1522^{+215}_{-150}$km s$^{-1}$, $L_{X_{bol}}=13.86\pm1.04\times 10^{44}$ erg s$^{-1}$, $kT=6.8^{+1.0}_{-0.6}$ keV. Mass estimate from weak lensing and X-ray are 3.7$\pm1.3 \times 10^{14}M_{\odot}$ and 4.35$\pm0.83\times 10^{14}M_{\odot}$, respectively \citep[see][for references]{2007MNRAS.382.1719K}.
 
 An important advantage of the AKARI observation is $L15$ filter, which corresponds to the restframe  8$\mu$m at z=0.81.
 With 15 (3) pointings, $L15$ reaches 66.5 (96.5)$\mu$Jy in deep (shallow) regions at 5$\sigma$.  
 Here flux is measured in 11'' aperture, and coverted to total flux using AKARI's IRC correction table (2009.5.1)\footnote{http://www.ir.isas.jaxa.jp/ASTRO-F/Observation/DataReduction/IRC/ApertureCorrection\_090501.html}.
 Cluster studies with the Spitzer are often performed in 24$\mu$m and thus needed a large extrapolation to estimate either $L_{8\mu m}$ or total infrared luminosity ($L_{TIR},8-1000\mu m$). 
  Note that we do not claim the $L_{8\mu m}$ is a better indicator of the total IR luminosity than other indicators \citep{2006ApJ...653.1129B,2007ApJ...666..870C,2009ApJ...692..556R}, but it is important that the AKARI can meausure redshifted $8\mu m$ flux directly in one of the filters.

 Thanks to the AKARI's wide field of view (10'$\times$10'), the total area coverage around the cluster is 200 arcmin$^2$, which cover larger area than previous cluster studies with the Spitzer, allowing us to study IR sources in the outskirts, where important galaxy evolution takes place \citep[e.g.,][]{2003MNRAS.346..601G}.   Previously, \citet{2008MNRAS.391.1758K} reported a high fraction of $L15$ sources in the intermediate density region in the cluster, suggesting a presence of environmental effect in the intermediate density environment.

 This same region was imaged with Suprime-Cam in $VRi'z'$ and has a good photometric redshift estimate \citep{2007MNRAS.382.1719K}. 
Used in this work are 54 $L15$-detected galaxies which are well identified with optical sources with $0.76\leq z_{photo}\leq 0.83$.

With the $L15$ filter covering the restframe 8$\mu$m, we simply convert the observed flux to 8$\mu$m monochromatic luminosity ($L_{8\mu m}$) using a standard cosmology. 
Completeness was measured by distributing artificial point sources with varying flux within the field and by examining what fraction of them was recovered as a function of input flux. Since we have deeper coverage at the center of the cluster, the completeness was measured separately in the central deep region and the outer regions of the field. More detail of the method is described in \citet{2008PASJ...60S.517W}.

 Once the flux is converted to luminosity and completeness is taken into account, it is straight forward to construct $L_{8\mu m}$ LFs, which we show in the squares in Fig.\ref{fig:8umlf}. Errors of the LFs are assumed to follow Poisson distribution.
Here, we take an angular distance of the most distant source  from the cluster center as a cluster radius ($R_{max}=6.2$Mpc).  We assumed $\frac{4}{3}\pi R_{\max}^3$ as the volume of the cluster to obtain galaxy density ($\phi$). This is only one of many ways to define a cluster volume, and thus, a caution must be taken to compare $absolute$ values of our LFs to other work such as \citet{Shim2010}. This cluster is elongated in angular direction \citep{2007MNRAS.382.1719K} , and thus, the volume might not be spherical. Yet, comparison of the $shape$ of the LFs is valid.

\subsection{LFs in the AKARI NEP Deep field}

Our field LFs are based on the AKARI NEP Deep field data. 
The AKARI performed deep imaging in the North Ecliptic Pole Region (NEP) from 2-24$\mu$m, with 4 pointings in each field over 0.4 deg$^2$ \citep{2006PASJ...58..673M,2007PASJ...59S.543M,2008PASJ...60S.517W}. 
 The 5 $\sigma$ sensitivity in the AKARI IR filters ($N2,N3,N4,S7,S9W,S11,L15,L18W$ and $L24$) are 14.2, 11.0, 8.0, 48, 58, 71, 117, 121 and 275$\mu$Jy \citep{2008PASJ...60S.517W}. 
Flux is measured in 3 pix radius aperture (=7''), then corrected to total flux. 

 A subregion of the NEP-Deep field (0.25 deg$^2$) has ancillary data from Subaru $BVRi'z'$\citep{2007AJ....133.2418I,2008PASJ...60S.517W}, CFHT $u'$(Serjeant et al. in prep.), KPNO2m/FLAMINGOs $J$ and $Ks$\citep{2007AJ....133.2418I},   GALEX $FUV$ and $NUV$ (Malkan et al. in prep.). For the optical identification of MIR sources, we adopt the likelihood ratio method  \citep{1992MNRAS.259..413S}.
Using these data, we estimate photometric redshift of $L15$ detected sources in the region with the {\ttfamily LePhare }
\citep{2006A&A...457..841I,2007A&A...476..137A}.
The measured errors on the photo-$z$ against 293 spec-z galaxies from Keck/DEIMOS (Takagi et al. in prep.) are $\frac{\Delta z}{1+z}$=0.036 at $z\leq0.8$. We have excluded those sources better fit with QSO templates from the LFs. 

 To construct field LFs, we have selected $L15$ sources at $0.65<z_{photoz}<0.9$. 
There remained 289 IR galaxies with a median redshift of 0.76. $L15$ flux is converted to $L_{8\mu m}$ using the photometric redshift of each galaxy. LFs are computed using the  1/$V_{\max}$ method. 
We used the SED templates \citep{2003MNRAS.338..555L} for $k$-corrections to obtain the maximum observable redshift from the flux limit.
 Completeness of the  $L15$ detection is corrected using \citet{Pearson15um}.
 This correction is 25\% at maximum, since we only use the sample where the completeness is greater than 80\%.

 The resulting field LFs are shown in the dotted line and triangles in Fig.\ref{fig:8umlf}.
 Errors of the LFs are computed using a 1000 Monte Caro simulation with varying $z$ and flux within their errors.
 These estimated errors are added to the Poisson errors in each LF bin in quadrature.

  We performed a detaild comparison of restframe 8$\mu$m LFs to those in the literature in \citet{GOTO_NEP_LF}.
 Briefly, there is an oder of difference between \citet{2007ApJ...660...97C} and \citet{2006MNRAS.370.1159B}, reflecting difficulty in estimating $L_{8\mu m}$ dominated by PAH emissions using Spitzer 24$\mu$m flux. Our field 8$\mu$m LF lies between \citet{2007ApJ...660...97C} and \citet{2006MNRAS.370.1159B}. Compared with these work, we have directly observed  restframe 8$\mu$m using the AKARI $L15$ filter, eliminating the uncertaintly in flux conversion based on SED models. More details and evolution of field IR LFs are described in \citet{GOTO_NEP_LF}.

\section{Results \& Discussion}\label{results}

\subsection{8$\mu m$ IR LFs}

In Fig.\ref{fig:8umlf}, we show restframe 8$\mu$m LFs of cluster RXJ1716.4$+$6708 in the squares, and LFs of the field region in the triangles. First of all, cluster LFs have by a factor of $\sim$700 higher density than the field LFs, reflecting the fact the galaxy clusters is indeed high density regions in terms of infrared sources.

Next, to compare the shape of the LFs, we normalized the cluster LF to match the field LFs at the faintest  end, and show in the dash-dotted line. 
 In contrast to the field LFs, which show flattening of the slope at log$L_{8\mu m}<10.8L_{\odot}$, the cluster LF maintains the steep slope in the range of $10.0L_{\odot}<\log L_{8\mu m}<10.6L_{\odot}$. 
 The difference is significant, considering the size of errors on each LF. 

 We fit a double-power law to both cluster and field LFs using the following formulae.

\begin{equation}
 \label{eqn:lumfunc2p}
 \Phi(L)dL/L^{*} = \Phi^{*}\bigg(\frac {L}{L^{*}}\bigg)^{1-\alpha}dL/L^{*}, ~~~ (L<L^{*})
\end{equation}

\begin{equation}
 \label{eqn:lumfunc2p2}
 \Phi(L)dL/L^{*} = \Phi^{*}\bigg(\frac {L}{L^{*}}\bigg)^{1-\beta}dL/L^{*}, ~~~  (L>L^{*})
\end{equation}

Free parameters are: $L^*$ (characteristic luminosity, $L_{\odot}$), $\phi^*$ (normalization, Mpc$^{-3}$), $\alpha$ and $\beta$ (faint and bright end slopes), respectively.
The best fit values for field and cluster LFs are summarised in Table\ref{tab:fit_parameters}
 and shown in the dotted lines in Fig.\ref{fig:8umlf}.

 The bright-end slopes are not very different, but $L^*$ of the cluster LF is smaller than the field by a factor of 2.4, and the faint-end tail of cluster LF is steeper than that of field LF.

To further examine the difference at the faint end of the LFs, we divide the cluster LF using the local galaxy density ($\Sigma_{5th}$) measured by \citet{2008MNRAS.391.1758K}. This density is based on the distance to the 5th nearest neighbor in the transverse direction using all the optical photo-z members, and thus, is a surface galaxy density. We separate LFs using similar criteria, $log \Sigma_{5th}\geq 2 $ (dense), $1.6\leq log \Sigma_{5th}<2$ (intermediate), and $log \Sigma_{5th}< 1.6$ (sparse), then plot LFs of each region in the stars, circles, and squares in Fig.\ref{fig:8umlf_density}. A fraction of the total volume of the cluster is assigned to each density group in inversely proportional to the sum of $\Sigma_{5th}^{3/2}$ of each group.

Interestingly, the faint-end slope becomes flatter and flatter with decreasing local galaxy density. This result is consistent with our comparison with the field in Fig.\ref{fig:8umlf}. In fact, the lowest density LF (squares) has a flat faint-end tail similar to that of the field LF.
 Since these LFs are based on the same data, 
 changes in the faint-end slope are not likely due to the errors in completeness correction nor calibration problems. The completeness of the deep and shallow regions of the cluster are measured separately. The changes in the slope is much larger than the maximum completeness correction of 25\%.
We also checked the cluster LFs as a function of cluster centric radius, to find no significant difference, perhaps due to the elongated morphology of this cluster. 
 At the same time,  assuming the same cluster volume, Fig.\ref{fig:8umlf_density} shows that a possible contamination from the field galaxies to cluster LFs is only $\sim$0.1\% in the dense region and $\sim$1\% even in the sparse region.


It is interesting that not just the change in the scale of the LFs, but there is a change in the $L^*$ and the faint-end slope ($\alpha$) of the LFs, resulting in the deficit in the  $10.2L_{\odot}<\log L_{8\mu m}<10.8L_{\odot}$ for cluster LFs. 
 One might imagine a change just in $L^*$ might explain the difference in Fig.\ref{fig:8umlf}. However, in Fig.\ref{fig:8umlf_density}, there clearly is a change in the slope as a function of $\Sigma_{5th}$.

However, interpretation is rather complicated; a shape of LF would not change if field galaxies infall into cluster uniformly without changing their star-formation activity. Although in cluster environment, a fraction of MIR luminous galaxies is smaller than field \citep{2008MNRAS.391.1758K}, uniform and instant quenching of star-formation activity of field galaxies can only shift a LF, but cannot account for a change in $L^*$ and $\alpha$ of the LFs.

Two important findings in this work are; (i) $L^*$ is smaller in the cluster. (ii) the faint-end slopes become steeper toward higher-density regions. To explain these changes in LFs, IR-luminous galaxies need to be preferentially reduced, with a relative increase of IR-faint galaxies. However,  an environmental-driven physical process such as the ram-pressure stripping or galaxy-merging would  quench star-formation not only in massive galaxies but in less massive galaxies as well, and thus is not able to explain the observed changes in LFs.
 
 On the other hand, it has been frequently observed that more massive galaxies formed earlier in the Universe.
This downsizing scenario also depends on the environment, in the sense that  galaxies with same mass are more evolved in higher density environments than galaxis in less dense environments \citep{2005ApJ...621..188G,2005MNRAS.362..268T,2008A&A...489..571T}. 
Statistically, a good correlation has been found between $L_{TIR}$ and stellar mass \citep{2007A&A...468...33E}.
Our finding of the relative lack of IR-luminous galaxies in the cluster environment may be consistent with the  downsizing scenario, where higher density regions have more evolved galaxies and lacks massive star-forming galaxies. In contrast, in lower density regions more massive galaxies are still star-forming.  However, since the data we have shown is in IR luminosity, to conclude on this, we need good stellar mass estimate based on deeper near-IR data.


Although a specific mechanism is unclear, the steep faint-end could also result from the enhanced star-formation in less massive galaxies. In the above scenario,  massive galaxies have already ceased their star-formation in the cluster, but less massive galaxies are still forming stars. These less massive galaxies may stop star-formation soon to join the faint-end of the red-sequence \citep{2007MNRAS.382.1719K}.

\subsection{Total IR LFs}


\begin{figure}
\begin{center}
\includegraphics[scale=0.6]{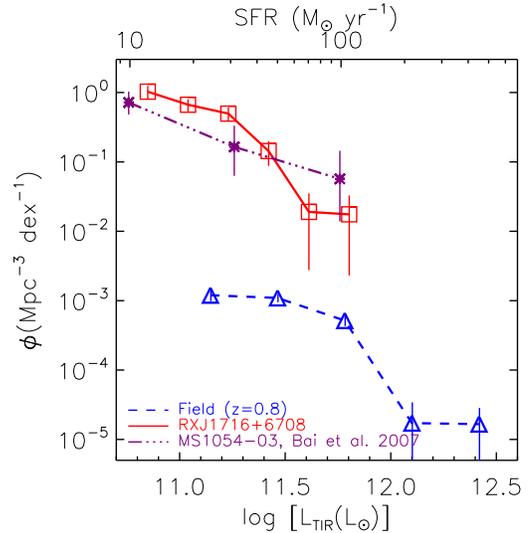}
\end{center}
\caption{
Total infrared LFs of cluster RXJ1716.4$+$6708 at z=0.81 in the solid line, and those of the AKARI NEP deep field in the dashed line.
Overplotted are the LFs of MS1054 from \citet{2007ApJ...664..181B}.
}\label{fig:tlf}
\end{figure}

To compare the $L_{8\mu m}$ LF in Fig.\ref{fig:8umlf} to those in the literature, we need to convert  $L_{8\mu m}$ to $L_{TIR}$.
We use the the following relation by \citet{2007ApJ...660...97C};

\begin{eqnarray}\label{8um_equation_caputi}
L_{TIR}=1.91\times (\nu L_{\nu_{8\mu m}})^{1.06} (\pm 55\%)
\end{eqnarray}

This is better tuned for a similar luminosity range used here than the original relation by \citet{2008A&A...479...83B}.
The conversion, however, has been the largest source of errors in estimating  $L_{TIR}$ from  $L_{8\mu m}$. \citet{2007ApJ...660...97C} report 55\% of dispersion around the relation. It should be kept in mind that the restframe $8\mu$m is sensitive to the star-formation activity, but at the same time, it is where the SED models have strongest discrepancies due to the complicated PAH emission lines \citep[see Fig.12 of ][]{2007ApJ...660...97C,GOTO_NEP_LF}.

The estimated  $L_{TIR}$ can be, then, converted to SFR using the following relation 
 for a Salpeter IMF, 
$\phi$ (m) 
$\propto m^{-2.35}$ between 
$0.1-100 M_{\odot}$  \citep{1998ARA&A..36..189K}.
\begin{eqnarray}
SFR (M_{\odot} yr^{-1}) =1.72 \times 10^{-10} L_{TIR} (L_{\odot}) 
\end{eqnarray}

In Fig.\ref{fig:tlf}, we show the $L_{TIR}$ LFs. Symbols are the same as Fig.\ref{fig:8umlf}.
In the top axis, we show corresponding SFR. 
Overplotted asterisks are cluster LF of MS1054 at z=0.83 with $\times$2 larger mass by \citet{2007ApJ...664..181B}, which show good agreement with our LFs of RXJ1716.4$+$6708 in the squares. 
 Note that \citet{2007ApJ...664..181B} covered only the central region of MS1054 due to the smaller field of view of the Spitzer.  The shape of their LF looks more similar to our LFs in the highest density bin in Fig.\ref{fig:8umlf_density}. A shift in scale is perhaps due to difference in esimating cluster volumes.

 A major difference of our work to that of \citet{2007ApJ...664..181B} is that they were not able to compare in detail on the shape of the LFs between field and cluster regions, due mainly to a smaller field coverage and larger errors on LFs. 
They had to fix the faint-end slope with a local value.
The largest source of errors is in converting Spitzer 24$\mu$m flux into 8$\mu m$. Both cluster and field LFs of this work use $L15$ filter, which measures restframe 8$\mu$m flux directly, eliminating the largest source of errors. In addition, both cluster and filed LFs are measured with an essentially same methodology, allowing us a fair comparison of LFs.


%
\section{Summary}\label{summary}

We constructed restframe 8$\mu$m LFs of a massive galaxy cluster (RXJ1716.4$+$6708) and a rarefied field region (the NEP deep field) at z$\sim$0.8 using essentially the same method and data from the AKARI telescope. 
AKARI's 15$\mu$m filter nicely covers restframe 8$\mu$m at z$\sim$0.8, and thus we do not need a large interpolation based on SED models. AKARI's wide field of view allows us to investigate variety of cluster environments with 2 orders of difference in local galaxy density.

We found that $L^*$ of the cluster 8$\mu$m LF is smaller than the field by a factor of 2.4, and the faint-end tail of cluster IR LFs become steeper and steeper with increasing local galaxy density. This difference cannot be explained by a simple infall of field galaxies into a cluster. 
 Physics that preferentially supresses IR luminous galaxes in higer density regions is needed to explain the observed results.

%
%
%

%
%
%
%

\section*{Acknowledgments}

We thank the anonymous referee for many insightful comments, which significantly improved the paper.
We are greateful to Masayuki Tanaka for useful discussion. We thank L.Bai for providing data for comparison.


T.G.,Y.K. and H.I. acknowledges financial support from the Japan Society for the Promotion of Science (JSPS) through JSPS Research Fellowships for Young Scientists.
MI was supported by the Korea Science and Engineering  Foundation(KOSEF) grant  No. 2009-0063616, funded by the Korea government(MEST)"
HML acknowledges the support from KASI through its cooperative fund in 2008. 




%

This research is based on the observations with AKARI, a JAXA project with the participation of ESA.

The authors wish to recognize and acknowledge the very significant cultural role and reverence that the summit of Mauna Kea has always had within the indigenous Hawaiian community.  We are most fortunate to have the opportunity to conduct observations from this sacred mountain.
%
%






\label{lastpage}

\end{document}